\documentstyle[12pt,epsf]{article}
\newcommand{\nc}{\newcommand}
\nc{\renc}{\renewcommand}

\nc{\etal}{\mbox{\it et al. }}
\nc{\ie}{{\it i.e.}}
\nc{\eg}{{\it e.g.}}

\renc{\thefootnote}{\arabic{footnote}}
\nc{\capt}[1]{{\bf Figure.} {\small\sl #1}}

\nc{\eqs}[2]{\mbox{Eqs.(\ref{#1},\,\ref{#2})}}
\nc{\eq}[1]{\mbox{Eq.(\ref{#1})}}

\nc{\figs}[2]{\mbox{Figs.(\ref{#1},\,\ref{#2})}}
\nc{\fig}[1]{\mbox{Fig.(\ref{#1})}}

\nc{\tag}[1]{\label{#1} \marginpar{{\footnotesize #1}}}
\nc{\mtag}[1]{\label{#1} \mbox{\marginpar{{\footnotesize #1}}}}
\renc{\baselinestretch}{1.2}
\newlength{\overeqskip}
\newlength{\undereqskip}
\nc{\be}[1]{\begin{equation} \mbox{$\label{#1}$}}
\nc{\bea}[1]{\begin{eqnarray} \mbox{$\label{#1}$}}
\nc{\Section}[2]{\section{#2}\label{#1}}
\nc{\Bibitem}[1]{\bibitem{#1}}
\nc{\Label}[1]{\label{#1}}

\nc{\eea}{\vspace{\undereqskip}\end{eqnarray}}
\nc{\ee}{\vspace{\undereqskip}\end{equation}}
\nc{\bdm}{\begin{displaymath}}
\nc{\edm}{\end{displaymath}}
\nc{\dpsty}{\displaystyle}
\nc{\bc}{\begin{center}}
\nc{\ec}{\end{center}}
\nc{\ba}{\begin{array}}
\nc{\ea}{\end{array}}
\nc{\bab}{\begin{abstract}}
\nc{\eab}{\end{abstract}}
\nc{\btab}{\begin{tabular}}
\nc{\etab}{\end{tabular}}
\nc{\bit}{\begin{itemize}}
\nc{\eit}{\end{itemize}}
\nc{\ben}{\begin{enumerate}}
\nc{\een}{\end{enumerate}}
\nc{\bfig}{\begin{figure}}
\nc{\efig}{\end{figure}}
\nc{\arreq}{&\!=\!&}
\nc{\arrmi}{&\!-\!&}
\nc{\arrpl}{&\!+\!&}
\nc{\arrap}{&\!\!\!\approx\!\!\!&}
\nc{\non}{\nonumber\\*}
\nc{\align}{\!\!\!\!\!\!\!\!&&}

\def\lsim{\; \raise0.3ex\hbox{$<$\kern-0.75em
      \raise-1.1ex\hbox{$\sim$}}\; }
\def\gsim{\; \raise0.3ex\hbox{$>$\kern-0.75em
      \raise-1.1ex\hbox{$\sim$}}\; }
\nc{\DOT}{\hspace{-0.08in}{\bf .}\hspace{0.1in}}
\nc{\Laada}{\hbox {$\sqcap$ \kern -1em $\sqcup$}}
\nc\loota{{\scriptstyle\sqcap\kern-0.55em\hbox{$\scriptstyle\sqcup$}}}
\nc\Loota{{\sqcap\kern-0.65em\hbox{$\sqcup$}}}
\nc\laada{\Loota}
\nc{\qed}{\hskip 3em \hbox{\BOX} \vskip 2ex}

\nc{\real}{{\rm I \! R}}
\nc{\Z}{{\sf Z \!\!\! Z}}
\nc{\complex}{{\rm C\!\!\! {\sf I}\,\,}}
\def\bigid{\leavevmode\hbox{\small1\kern-3.8pt\normalsize1}}
\def\id{\leavevmode\hbox{\small1\kern-3.3pt\normalsize1}}
\nc{\slask}{\!\!\!/}
\nc{\bis}{{\prime\prime}}
\nc{\pa}{\partial}
\nc{\na}{\nabla}
\nc{\ra}{\rangle}
\nc{\la}{\langle}
\nc{\goto}{\rightarrow}
\nc{\swap}{\leftrightarrow}

\nc{\EE}[1]{ \mbox{$\cdot10^{#1}$} }
\nc{\abs}[1]{\left|#1\right|}
\nc{\at}[2]{\left.#1\right|_{#2}}
\nc{\norm}[1]{\|#1\|}
\nc{\abscut}[2]{\Abs{#1}_{\scriptscriptstyle#2}}
\nc{\vek}[1]{{\rm\bf #1}}
\nc{\integral}[2]{\int\limits_{#1}^{#2}}
\nc{\inv}[1]{\frac{1}{#1}}
\nc{\dd}[2]{{{\partial #1}\over{\partial #2}}}
\nc{\ddd}[2]{{{{\partial}^2 #1}\over{\partial {#2}^2}}}
\nc{\dddd}[3]{{{{\partial}^2 #1}\over
	{\partial #2 \partial #3}}}
\nc{\dder}[2]{{{d #1}\over{d #2}}}
\nc{\ddder}[2]{{{d^2 #1}\over{d {#2}^2}}}
\nc{\dddder}[3]{{d^2 #1}\over
	{d #2 d #3}}
\nc{\dx}[1]{d\,^{#1}x}
\nc{\dy}[1]{d\,^{#1}y}
\nc{\dz}[1]{d\,^{#1}z}
\nc{\dl}[1]{\frac{d\,^{#1}l}{(2\pi)^{#1}}}
\nc{\dk}[1]{\frac{d\,^{#1}k}{(2\pi)^{#1}}}
\nc{\dq}[1]{\frac{d\,^{#1}q}{(2\pi)^{#1}}}

\nc{\cc}{\mbox{$c.c.$ }}
\nc{\hc}{\mbox{$h.c.$ }}
\nc{\cf}{cf.\ }
\nc{\erfc}{{\rm erfc}}
\nc{\Tr}{{\rm Tr\,}}
\nc{\tr}{{\rm tr\,}}
\nc{\pol}{{\rm pol}}
\nc{\sign}{{\rm sign}}
\nc{\bfT}{{\bf T }}

\nc{\cA}{{\cal A}}
\nc{\cB}{{\cal B}}
\nc{\cD}{{\cal D}}
\nc{\cE}{{\cal E}}
\nc{\cG}{{\cal G}}
\nc{\cH}{{\cal H}}
\nc{\cL}{{\cal L}}
\nc{\cO}{{\cal O}}
\nc{\cT}{{\cal T}}
\nc{\rvac}[1]{|{\cal O}#1\rangle}
\nc{\lvac}[1]{\langle{\cal O}#1|}
\nc{\rvacb}[1]{|{\cal O}_\beta #1\rangle}
\nc{\lvacb}[1]{\langle{\cal O}_\beta #1 |}
\nc{\bb}{\bar{\beta}}
\nc{\bt}{\tilde{\beta}}
\nc{\ctH}{\tilde{\cal H}}
\nc{\chH}{\hat{\cal H}}
\nc{\1}{\aa}
\nc{\2}{\"{a}}
\nc{\3}{\"{o}}
\nc{\4}{\AA}
\nc{\5}{\"{A}}
\nc{\6}{\"{O}}
\nc{\al}{\alpha}
\nc{\g}{\gamma}
\nc{\Del}{\Delta}
\nc{\e}{\epsilon}
\nc{\eps}{\epsilon}
\nc{\lam}{\lambda}
\nc{\om}{\omega}
\nc{\Om}{\Omega}
\nc{\ve}{\varepsilon}
\nc{\mn}{{\mu\nu}}
\nc{\k}{\kappa}
\nc{\vp}{\varphi}

\nc{\advp}[3]{{\it  Adv.\ in\ Phys.\ }{{\bf #1} {(#2)} {#3}}}
\nc{\annp}[3]{{\it  Ann.\ Phys.\ (N.Y.)\ }{{\bf #1} {(#2)} {#3}}}
\nc{\apl}[3]{{\it  Appl. Phys. Lett. }{{\bf #1} {(#2)} {#3}}}
\nc{\apj}[3]{{\it  Ap.\ J.\ }{{\bf #1} {(#2)} {#3}}}
\nc{\apjl}[3]{{\it  Ap.\ J.\ Lett.\ }{{\bf #1} {(#2)} {#3}}}
\nc{\app}[3]{{\it Astropart.\ Phys.\ }{{\bf #1} {(#2)} {#3}}}
\nc{\cmp}[3]{{\it  Comm.\ Math.\ Phys.\ }{{ \bf #1} {(#2)} {#3}}}
\nc{\cqg}[3]{{\it  Class.\ Quant.\ Grav.\ }{{\bf #1} {(#2)} {#3}}}
\nc{\epl}[3]{{\it  Europhys.\ Lett.\ }{{\bf #1} {(#2)} {#3}}}
\nc{\ijmp}[3]{{\it Int.\ J.\ Mod.\ Phys.\ }{{\bf #1} {(#2)} {#3}}}
\nc{\ijtp}[3]{{\it Int.\ J.\ Theor.\ Phys.\ }{{\bf #1} {(#2)} {#3}}}
\nc{\jmp}[3]{{\it  J.\ Math.\ Phys.\ }{{ \bf #1} {(#2)} {#3}}}
\nc{\jpa}[3]{{\it  J.\ Phys.\ A\ }{{\bf #1} {(#2)} {#3}}}
\nc{\jpc}[3]{{\it  J.\ Phys.\ C\ }{{\bf #1} {(#2)} {#3}}}
\nc{\jap}[3]{{\it J.\ Appl.\ Phys.\ }{{\bf #1} {(#2)} {#3}}}
\nc{\jpsj}[3]{{\it J.\ Phys.\ Soc.\ Japan\ }{{\bf #1} {(#2)} {#3}}}
\nc{\lmp}[3]{{\it Lett.\ Math.\ Phys.\ }{{\bf #1} {(#2)} {#3}}}
\nc{\mpl}[3]{{\it  Mod.\ Phys.\ Lett.\ }{{\bf #1} {(#2)} {#3}}}
\nc{\ncim}[3]{{\it  Nuov.\ Cim.\ }{{\bf #1} {(#2)} {#3}}}
\nc{\np}[3]{{\it  Nucl.\ Phys.\ }{{\bf #1} {(#2)} {#3}}}
\nc{\pr}[3]{{\it Phys.\ Rev.\ }{{\bf #1} {(#2)} {#3}}}
\nc{\pra}[3]{{\it  Phys.\ Rev.\ A\ }{{\bf #1} {(#2)} {#3}}}
\nc{\prb}[3]{{\it  Phys.\ Rev.\ B\ }{{{\bf #1} {(#2)} {#3}}}}
\nc{\prc}[3]{{\it  Phys.\ Rev.\ C\ }{{\bf #1} {(#2)} {#3}}}
\nc{\prd}[3]{{\it  Phys.\ Rev.\ D\ }{{\bf #1} {(#2)} {#3}}}
\nc{\prl}[3]{{\it Phys\ Rev.\ Lett.\ }{{\bf #1} {(#2)} {#3}}}
\nc{\pl}[3]{{\it  Phys.\ Lett.\ }{{\bf #1} {(#2)} {#3}}}
\nc{\prep}[3]{{\it Phys\. Rep.\ }{{\bf #1} {(#2)} {#3}}}
\nc{\prsl}[3]{{\it Proc.\ R.\ Soc.\ London\ }{{\bf #1} {(#2)} {#3}}}
\nc{\ptp}[3]{{\it  Prog.\ Theor.\ Phys.\ }{{\bf #1} {(#2)} {#3}}}
\nc{\ptps}[3]{{\it  Prog\ Theor.\ Phys.\ suppl.\ }{{\bf #1} {(#2)} {#3}}}
\nc{\physa}[3]{{\it  Physica\ A\ }{{\bf #1} {(#2)} {#3}}}
\nc{\physb}[3]{{\it  Physica\ B\ }{{\bf #1} {(#2)} {#3}}}
\nc{\phys}[3]{{\it Physica\ }{{\bf #1} {(#2)} {#3}}}
\nc{\rmp}[3]{{\it  Rev.\ Mod.\ Phys.\ }{{\bf #1} {(#2)} {#3}}}
\nc{\rpp}[3]{{\it Rep.\ Prog.\ Phys.\ }{{\bf #1} {(#2)} {#3}}}
\nc{\sjnp}[3]{{\it Sov.\ J.\ Nucl.\ Phys.\ }{{\bf #1} {(#2)} {#3}}}
\nc{\spjetp}[3]{{\it Sov.\ Phys.\ JETP\ }{{\bf #1} {(#2)} {#3}}}
\nc{\yf}[3]{{\it Yad.\ Fiz.\ }{{\bf #1} {(#2)} {#3}}}
\nc{\zetp}[3]{{\it Zh.\ Eksp.\ Teor.\ Fiz.\  }{{\bf #1}  {(#2)} {#3}}}
\nc{\zp}[3]{{\it Z.\ Phys.\ }{{\bf #1} {(#2)} {#3}}}
\nc{\ibid}[3]{{\sl ibid.\ }{{\bf #1} {#2} {#3}}}
\nc{\rf}[1]{(\ref{#1})}
\nc{\nn}{\nonumber \\*}
\nc{\bfB}{\bf{B}}
\nc{\bfv}{\bf{v}}
\nc{\bfx}{\bf{x}}
\nc{\bfy}{\bf{y}}
\nc{\vx}{\vec{x}}
\nc{\vy}{\vec{y}}
\nc{\oB}{\overline{B}}
\nc{\oI}{\overline{I}}
\nc{\oR}{\overline{R}}
\nc{\rar}{\rightarrow}
\nc{\ti}{\times}
\nc{\slsh}{\hskip-5pt/}
\nc{\sm}{Standard~Model~}
\nc{\MP}{M_{\rm Pl}}
\nc{\tp}{t_{\rm Pl}}
\nc{\ave}{\bar{E}}

\renc{\min}{p_{\rm min}}
\renc{\max}{p_{\rm max}}
\nc{\pmin}{p_{\rm min}}
\nc{\pmax}{p_{\rm max}}
\nc{\fo}{f_0}
\nc{\foi}{f_{0,i}\,}
\nc{\fop}{f_0^P}
\nc{\fou}{f_0^U}
\def\sepand{\rule{14cm}{0pt}\and}
\nc{\eff}{{\rm eff}}
\nc{\MT}{M_{\rm T}}
\nc{\ML}{M_{\rm L}}
\nc{\kk}{\vek{k}}
\nc{\pp}{{\rm p}}
\begin{document}

{\title{{\hfill {{\small NORDITA-93/80\ P\\
	 \hfill hep-ph/9312265 }}\vskip 1truecm}
{\bf On the Non--equilibrium  Early Universe}}


\author{
{\sc Per Elmfors$^{1}$
and Kari Enqvist$^{2}$ } \\
{\sl Nordita, Blegdamsvej 17} \\
{\sl DK-2100 Copenhagen \O, Denmark}\\
and\\
{\sc Iiro Vilja$^{3}$ }\\
{\sl Department of Physics,
University of Turku} \\
{\sl FIN-20500 Turku, Finland} \\
\sepand
}
\maketitle}
\vspace{2cm}
\begin{abstract}
\noindent
We study non--equilibrium ensemble corrections to particle masses and the
effective potential in the early universe, using a uniform  momentum
distribution as an example. The resulting thermalization
temperature is computed assuming \sm degrees of freedom, and it is always found
  to be below $5\times 10^{14}$ GeV, implying that GUT phase transitions
typically take place out of equilibrium. For the abelian Higgs model we find
that the phase transition is of second order if it occurs before equilibration,
in contrast to the  first order equilibrium phase transition.
\end{abstract}
\vfill
\footnoterule
{\small $^1$elmfors@nordita.dk; $^2$enqvist@nbivax.nbi.dk;
$^3$vilja@sara.cc.utu.fi}
\thispagestyle{empty}
\newpage
\setcounter{page}{1}
\Section{intro}{Introduction}
The very early universe is usually assumed to have been in a state of thermal
 equilibrium, or in a state where entropy is maximal.
It was, however, estimated already in \cite{EllisS80} that GUT interactions
would not be rapid enough to maintain equilibrium above $T\simeq 3\ti 10^{15}$
 GeV, and in \cite{EnqvistS93}
it was shown by direct computation  in the \sm that quarks and gluons are no
 longer in
chemical equilibrium when $T\gsim 3\ti 10^{14}$ GeV. Going back in time, the
 expansion rate of the universe becomes so fast that particle reaction
rates can no longer keep up with it. Unless
for some reason interactions at the Planck scale can act over distances
larger than the causal domain, particles inside different horizons would
be completely uncorrelated. If the interaction is weak, they would also remain
uncorrelated for some time
 even after having entered the same causal domain.

Therefore, there seems to be no compelling reason for the universe to have been
in a state of maximal entropy at the very earliest times. It is of course
conceivable that gravitational effects (e.g. black hole evaporation) near the
Planck time $\tp$ could have resulted in a maximum entropy state for
non--gravitational quanta independently within each causal domain, but
naturally
this remains to be shown. Moreover, there are finite size effects in the
distribution function for a maximal entropy state, when the causal volume, and
the number of states in it,  is very
small. In the absence
of a theory of quantum gravity, which would give a dynamical description of
the universe also at the Planck scale, we explore here the consequences of the
assumption that the initial condition
for the classical universe, described in terms of the Einstein equations and
quantum matter fields, was not in a conventional state of maximal entropy.

Particles traversing a gas of quanta   experience forward scattering, which
modifies their free dispersion. This remains true also in the case of a
non--equilibrium system.
Forward scattering  manifests itself, for example, as an ensemble
correction to the masses of particles. The effective potentials for order
parameters, such as the Higgs field,   also receive ensemble corrections,
which in general differ from thermal corrections. Phase
transitions still exist in a non--equilibrium universe, but their dynamics
may not resemble the thermal case.
Because the forward scattering rate is usually larger than the
non--forward rate,  non--equilibrium ensemble corrections are present for
some time as the universe  keeps on expanding. Afterwards non--forward
reaction rates begin to exceed the expansion rate, and the system
thermalizes.

In the present paper we discuss  particle propagation in a non--equilibrium
background and study the nature of
phase transitions in the very early universe. We use a simple example
to demonstrate some differences between equilibrium and non--equilibrium
universes, and we show that thermalization is in general expected to
take place somewhat below typical GUT energy scales.
%
\Section{noneq}{Non--equilibrium dynamics}
Let us begin by discussing the formulation of a quantum field theory for a
non--equilibrium system, and show  under which assumptions we can replace
the thermal distribution functions in the real--time propagators by a more
general distribution. (For a treatment of quantum fields in mixed states see
e.g. \cite{joku}.)
We shall assume that the probability of finding one particle in a state
labeled by $\kk$ is given by $p_\kk$ and that the particles are uncorrelated.
Then the probability of finding $n$ particles in that state
is $p_\kk^n$ if they are bosons, and zero for $n\geq 2$ if they are fermions.
The density matrix for this kind of state can be constructed
explicitly using the number operator for each state $n_\kk=a_\kk^\dagger
a_\kk$. The probability depends  exponentially on $n_\kk$. Assuming
different states to be uncorrelated we are led to a density matrix which
factorizes as a product over all states. We can write it as
\be{densmatr}
	\rho=\frac{\exp\left[-\int\dk{3} h_\kk a^\dagger_\kk a_\kk\right]}
	{\Tr\exp\left[-\int\dk{3} h_\kk a^\dagger_\kk a_\kk\right]}\ ,
\ee
where $h_\kk$ is a dimensionless function.
This density matrix shares many nice properties with the equilibrium density
matrix for a non--interacting system used in standard
real--time perturbation theory, where
the particular choice of $h(\kk)$ is a linear combination of the energy
and conserved charges in the momentum state $\vek{k}$. Also in that case
multiparticle correlations in the
initial state are neglected.

Given the density matrix in \eq{densmatr}, we would like to derive a
perturbation theory for the statistical average of Green functions. We cannot
use the imaginary time formalism since there is no relation between
the density
matrix and the time evolution operator which is of essential
importance in the equilibrium
 case. There is, however, the real--time formalism
of Thermo Field Dynamics \cite{UmezawaMT82} which suites our
purposes. A
thermal vacuum for the density matrix in \eq{densmatr} can easily be
 constructed as
\be{thermvac}
	 \rvac{}=\exp\left[\int\dk{3} \vartheta(\kk)
	(a_\kk^\dagger\tilde{a}_\kk^\dagger-a_\kk\tilde{a}_\kk)\right]
	|0,\tilde{0}\ra\ ,
\ee
where $\sinh^2\vartheta(\kk)=(e^{h(\kk)}-1)^{-1}$ for bosons and
$\sin^2\vartheta
(\kk)=(e^{h(\kk)}+1)^{-1}$ for fermions. The function $h(\kk)$ is in principle
arbitrary but the number expectation value must be positive. It need not be
the same for bosons and fermions and it may depend on the spin and the charge
of the state. Since we are  interested in a neutral universe we choose
$h(\kk)$ not to depend on the charge, i.e. $h(\kk)$ is the same function for
particles and antiparticles.
For simplicity we also take the same spin independent $h(\kk)$ for bosons
and fermions. The distribution functions are then
\be{distrfct}
	f_{B(F)}(\vek{k})=\frac{1}
	{e^{h(\vek{k})}\;\raise.3ex\hbox{--\kern-.8em\lower.3ex
	\hbox{$_{(+)}$}}\; 1}\ .
\ee

It is straightforward to find the (11)--component of the scalar, fermion and
gauge boson
 propagators
for non--interacting fields:
\bea{1}
	D_B(k)&=&{i\over k^2-m^2+i\epsilon}+
	2\pi\delta(k^2-m^2)f_B(k)\ , \nn
	D_F(k)&=&{i(k\slsh +m)\over k^2-m^2+i\epsilon}-
	2\pi\delta(k^2-m^2)(k\slsh +m)f_F(k)\ ,\nn
	D_{GB}(k)&=&-g_\mn\left({i\over k^2+i\epsilon}+
	2\pi\delta(k^2)f_B(k)\right)\ .
\eea
These propagators are, of course, the same as in equilibrium except
that the one--particle distribution functions are different.
%
\Section{enscorr}{Ensemble corrections}
The effective action is in general very difficult to calculate in an expanding
universe where both the fields and the thermal state depend on
time.\footnote{See however \cite{SemW} and references therein.}
 We shall
consider a universe where the expansion is slow as compared with the
oscillation frequency of the modes. This means that in order for the ensemble
corrections of the masses to exist, we should require that the
Hubble rate satisfies $H\ll m$.

By virtue of the uncertainty principle, waves with $p\lsim H$ do
not fit into a single horizon. If such particle modes existed in the early
universe,
for an observer inside a given horizon they would appear as background fields
rather than quanta. The size
of the horizon thus provides an infrared cut--off for particle momenta.
We   neglect this cut--off, which
gives rise only to a small correction whenever the average energy
per particle $\ave$ satisfies $\ave\gg H$.

Using
\eq{1}, one easily obtains the ensemble masses for gauge bosons,
(chiral) fermions $\psi^{L(R)}$ in representation $R_{L(R)}$ and scalars $\phi$
 in
representation $\Gamma$ with a Yukawa coupling $g_Y \bar\psi_m^L\Gamma_{mn}^i
\psi_n^R\phi_i$ and a scalar potential $-\mu^2{\rm Tr}\;\phi^\dagger\phi +
\lambda {\rm Tr}\;(\phi^\dagger\phi)^2$
in an SU(N)
 theory (we drop possible ${\rm Tr}\;\phi^3$ terms for simplicity).
In the limit when $\ave$ is much larger than the   masses of
the particles, at one loop and in the Feynman
gauge we find the following static masses:
\bea{4}
	m^2_{F,L(R)}\arreq\biggl[2 g^2C(R_{L(R)}) +
	    |g_Y|^2 C_{L(R)}'(\Gamma)\biggr]\biggl[\la k^{-1}\ra_F
            +\la k^{-1}\ra_B\biggr]\ , \nn
	M^2_{L}\arreq 4g^2\biggl[ 2T(R)\la k^{-1}\ra_F+
		\bigl(N+T(\Gamma)\bigr)\la k^{-1} \ra_B\biggr]\ ,\nn
	m^2_s\arreq -\mu^2 + \left[3 g^2C(\Gamma) +2\lambda \left(1 + {\rm
dim}(\Gamma)\right)\right]
\la k^{-1} \ra_B +
	6 |g_Y|^2\tilde C(\Gamma)
             \la k^{-1} \ra_F\ ,
\eea
where $C(R_{L(R)})$ and $C(\Gamma)$ are the quadratic Casimir invariants of the
 representations $R_{L(R)}$ and $\Gamma$, respectively,
$T(R)$ ($T(\Gamma)$) is the index of the representation $R$ ($\Gamma$) and
$C_{L(R)}'(\Gamma)$ and $\tilde C(\Gamma)$ are defined by $(\Gamma ^i
\Gamma^{i\, \dagger})_{nm} = C_L'(\Gamma)\delta_{mn}$,
$(\Gamma^{i\, \dagger}\Gamma^i)_{mn} = C_R'(\Gamma)\delta_{mn}$ and
${\rm tr}(\Gamma^i\Gamma^{j\, \dagger}) = \tilde C(\Gamma) \delta^{ij}$. The
longitudinal
 mass of the gauge boson is denoted by $M_{L}$ and the transverse mass,
$M_{T}$, is zero. We have used the notation
\be{average}
	\la F(k) \ra_i\equiv\int\dk{3} F(k) f_i(k)\
\ee
for the average of a function $F(k)$.

Also, a general expression for the thermal part of the effective
potential can be given. It reads,
besides an additive $\phi$-independent constant,
\be{pot}
V_{eff}(\phi) = \sum_{i} \int\dk{3}\ \int_{\infty}^{\omega_k(m_i)}
d\omega\, f_i(\omega)\ ,
\ee
where $\omega_k(m_i) = \sqrt{k^2 + m_i^2(\phi)}$, $m_i(\phi)$ is the
eigenvalue of the mass matrix corresponding to the  $i$-th
degree of freedom in  a $\phi$--dependent background and the sum is over all
bosonic and fermionic degrees of freedom.
%
\Section{thermo}{Evolution of the initial distribution}
The assumption about the universe as a collection of statistically independent,
massless, one--particle
states simplifies the equation of state. It is also a natural approximation for
small coupling constants since then the typical momentum is much larger than
the ensemble induced mass. In a flat Robertson--Walker universe the Hubble
parameter is simply
given by
\be{hubble}
	H=\inv{2t}=\inv{r_H}=
	\left(\frac{8\pi\cE_{\rm tot}}{3\MP^2}\right)^{1/2}
\ee
also in the non--equilibrium case.
In the absence of particle--particle correlations
we may immediately write down the equation of state as $\pp=\cE /3$
by virtue of the definition of the energy--momentum tensor.
The energy and number densities read as
$\cE_i =\la|{\bf k}|\ra_i \ , ~~ n_i=\la 1\ra_i\ ; ~~~(i=F,B)~,
$
with average energy $\ave =\cE_{tot}/n$  also depending on the
distribution.

It is useful to consider radiation in
thermal equilibrium and go back in time to the Planck scale in order to see
what the typical values of thermodynamical quantities are, even if we do not
believe that the universe was in equilibrium at that time. Using the notation
$\alpha=90/8\pi^3g_*\simeq 0.0034$ and $\tau=t/\tp$ we find the
following expressions
\bea{eqexpr}
	r_H\,\MP&=& \MP/H= 2\tau\ ,\nn
	N_H&=&n r_H^3= 2^{3/2}h_*\zeta(3)\alpha^{3/4}\tau^{3/2}/\pi^2
	\simeq
	0.46\tau^{3/2}\ ,\nn
	 \cE r_H^3/\MP&=& 3\tau/4\pi
	\simeq
	0.24\tau\ ,\nn
	 T/\MP&=& \alpha^{1/4}
	/(2\tau)^{1/2}\simeq
	0.171\tau^{-1/2}\ ,
\eea
where the \sm values $g_*=g_B+\frac{7}{8}g_F=106.75$, $h_*=g_B+\frac{3}{4}
g_F=95.5$, $g_B=28$, $g_F=90$, are used for reference purposes.

Our simple  picture is likely to be distorted by gravitational
effects near the Planck scale. In a universe dominated by curvature effects
there is no unambiguous definition of the particle vacuum, since it depends
on the inertial frame \cite{birrell}.
For example, accelerated observers are immersed in a bath of
thermal radiation. The particle creation rate
goes, however, to zero exponentially
when $k\gg H$, so that such modes can be described by an adiabatic
vacuum. Moreover, in an isotropic universe comoving observes have the
same vacuum. Therefore, if we assume isotropy, homogeneity
 and neglect curvature effects,
the description of the early universe presented here should be  sensible when
$\ave\ll\MP$.
Consistency requires  that we also neglect the coupling of quantum fields
directly to curvature. Our range of validity is then such that $H$ should
be small compared with any other scale in the problem. Moreover, the
gravitational field
should  be classical so that we must require that $H\ll\MP$.

To be more specific, let us now consider an example for the initial
particle distribution. To reiterate our starting point, let us assume
that  causal domains are uncorrelated and
there was no (gravitational) mechanism which would have maximized
entropy within each horizon. We may then envisage a simple idealization:
the particle energy within a  causal domain was  randomly
distributed up to a maximum energy. We shall refer to this as the
uniform distribution,  and define
\be{10}
	~~f_i(p)=\foi\theta(\pmax(t)-p)\ ,
\ee
with
$f_{0,B(F)}=( e^{h_0}\;\raise.3ex\hbox{--\kern-.8em
\lower.3ex\hbox{$_{(+)}$}}\; 1)^{-1}$. The time
dependence of  $\pmax(t)$ is simply determined
by the redshift from the expansion of the universe. We then have for
each degree of freedom
\be{12}
	\cE_i =  {1\over 8\pi^2}\foi\max^4\ ,~~
	n_i = {1\over 6\pi^2}\foi\max^3\ ,~~\ave= \frac 34\max\ ,
	\quad \la \inv{k}\ra_i = \inv{4\pi^2}\foi\pmax^2\ ,
\ee
so that the average energy  is independent of $\foi$. The total energy density
is
given by $\cE_{tot}=f_*\max^4(t)/8\pi^2$ with $f_*=\sum
f_{0,i}=g_Bf_{0,B}+g_Ff_{0,F}$.

The initial condition is given in terms of the number of particles and
the total energy within a horizon. At $t=\tp$ we would expect from \eq{eqexpr}
that $\ave\sim
\MP$ and $N_H\sim 1$. The relation $H(\tp)=1/2\tp$ determines $\cE_{tot}(\tp)$
in a flat universe to be $3\MP^4/32\pi$ so that $\pmax(\tp)$ is in fact
directly related to $f_*$.
Thus $N_H(\tp)=0.26f_*^{1/4}$ and if
$N_H(\tp)\sim N_{H,SM}(\tp)$, then $\foi\sim 0.1$.
Let us again emphasize that although our initial conditions are given
at the Planck scale, we do not claim that one could extrapolate physics
all the way up to that scale. Rather, \eq{10} should be considered as
valid only when $\ave\ll\MP$.
If there are no intervening phase transitions, such as inflation, which convert
background fields to radiation, one can directly compute the thermalization
temperature from the initial condition.
%
\Section{relax}{Thermalization}
Let us, in this Section, consider the  Standard Model and assume that there are
no phase transition before thermalization. The range of validity of the
relations in \eq{12} depend on the thermalization
rate $\g$, which for scalars is related to the imaginary part of the two--point
function via $\g={\rm Im\;}\Gamma^{(2)}(\omega,{\bf p})/\omega$.
Here $\omega=\omega ({\bf p})$ is the solution to the dispersion relations.
At one
loop the imaginary part of the Higgs propagator receives
contributions from decay
and absorption (emission) processes. Absorption (emission) is always
proportional
to the difference between the distribution functions of the two
external  particles
in the final and initial state \cite{ElmforsEV93b}.
Thus in the uniform case,
where the distributions do not depend on the momentum, this difference is zero.
Other
distributions, with a definite
momentum dependence, could result in a non--zero imaginary part, but for
the present example we have to go to next order.

In the \sm at high $T$, where in the case of light Higgs we have $m^2_H(T)
\equiv C_H^{eq}T^2\simeq 0.24\;T^2$, kinematics forbids decay into quarks.
Leptonic channels are allowed,
but for them the thermalization rate, which depends on the Yukawa coupling
squared, is very small.

A similar argument holds also for gauge boson thermalization, which in the
thermal case has been computed for QCD in \cite{braaten}. That calculation
cannot directly be taken over to non--equilibrium case because of the
infra--red singularities, but here too kinematics simply
prevents an imaginary part from appearing at one loop in the case of
uniform distribution.

For fermions the one--loop kinematical suppression
is not complete but enters in the
form $f_{0,B}-f_{0,F}=2/({\rm exp}(2h_0)-1)$, which is small if $f_{0,B}$ is
small. For
our purposes it suffices to observe that although fermionic equilibration
is a more involved issue here, at least bosonic equilibration does not
take place at one loop due to kinematical constraint, so that the
system as a whole is not equilibrated.

At two loop kinematical constraints no longer exist.
The leading purely bosonic contribution to the Higgs thermalization rate
in the \sm has been
estimated in \cite{ElmforsEV93b} from the gauge boson loops;
the result can be taken over to the uniform case by
replacing $T^2\to 3f_{0,B}\max^2(t)/\pi^2$, which yields
\be{16}
	\g\simeq {9g^4\over 256\pi^3}{f_{0,B}\max^2(t)\over m_H}\ ,
\ee
where, neglecting light fermions,
$m_H^2\simeq (9g^2f_{0,B} + 24 g_{top}^2f_{0,F})\pmax^2(t)/16\pi^2\equiv
C_H\max^2(t)$
and we have taken $p_0=m_H,~\vek p =0$. This expression is valid well
above a possible phase transition.

Thermalization begins when $\g\simeq H=(2t_{th})^{-1}$, and
assuming no intervening phase transitions,
one may determine $\max(t_{th})$ through
\be{161}
   \max(t_{th})=3.1\times 10^{-4}f_{0,B}(C_Hf_*)^{-1/2}\MP\ .
\ee
Here we have set $g^2\simeq 0.3$, in anticipation that the relevant
energy scales are GUT scales. We also fix $m_{top}=135$ GeV for definiteness.
We note that $1.4\times 10^{-4}\le\max/\MP\le 4.4\times 10^{-4}$ as
$\infty>h_0>0$. The thermalization time is given by
$t_{th}=1.6\times10^7C_Hf_*^{1/2}f_{0,B}^{-2}\MP^{-1}$.
If we assume
that thermalization is instantaneous, we find the resulting temperature to be
\be{17}
	T_{th}= 1.4\times 10^{-4}(C^2_Hf_*g_*)^{-1/4}f_{0,B}\MP~.
\ee
For comparison,
 assuming thermal distributions would yield $T_{th}^{eq}=5\times
10^{14}$ GeV, a result which is also consistent with the computation in
\cite{EnqvistS93}. Thus, if $T_{th}$ as computed from \eq{17} is higher than
$T_{th}^{eq}$, it would be inconsistent to assume
relaxation to full equilibrium
at $T_{th}$, but if $T_{th}\leq T_{th}^{eq}$ thermalization takes place at
$T_{th}$ and is probably rather fast. In the Figure we display the
thermalization temperature
against the number of particles within a horizon at $\tp$.
\bfig[t]
\epsfxsize=16cm
\epsfbox{kuvaout.ps}
\baselineskip 13pt
\capt{Solid line: Thermalization temperature as a function of number of
particles within
a horizon volume. The \sm particle
content is assumed.   Dashed line: The limit on the scalar mass, above
which phase transition occurs in non--equilibrium environment,
is also shown. The
slope is universal while the horizontal line is for the \sm
degrees of freedom only.}
\efig

We have checked that in the range depicted in the Figure, the number of
particles at thermalization is always large, and that $H(T_{th})\ll \MP$.
Likewise, $m_H\gg H$ provided $\max/\MP\ll 0.06\; (0.08)$ for
$f_{0,B}\to \infty\; (0)$, which according to \eq{161} is always true at
$t_{th}$.
It therefore
appears that relaxation to equilibrium in the case of uniform initial
distribution takes place at, or somewhat below, typical GUT temperatures
which are much less than $\MP$.
This justifies our neglecting of the space--time curvature. Above the
thermalization temperature we  have $m_H\gg H>\gamma$ so that forward
scattering  dominates over the non--forward rate and the particle excitations
are relatively stable.
%
\Section{phtr}{Phase transitions}
 The notion of an equilibrium  GUT phase transition is
questionable and must be examined case by case. However,
as long as the non--equilibrium effective potential changes slowly as
compared to the field
oscillations, it is still meaningful to discuss the position of the minimum of
the effective potential and how it changes as the universe expands.
In this Section we use the uniform distribution though the actual distribution
may change close to the  transition temperature.

The ensemble correction to masses are similar in and out of
equilibrium, as we have seen in Section \ref{enscorr}. We
can only expect essential differences for quantities which depend in a crucial
way on the particular form of the distribution function. The first order nature
of the phase
transition in the abelian Higgs model can be traced to the cubic term in
$\phi$ which is generated by the infra--red divergent
part of the Bose--Einstein
distribution function. Therefore, for a uniform distribution
we do not expect any first order phase
transition. We can also argue that it is the massless states that give rise to
the infra--red singularity  since it
costs little energy to produce them and they contribute a lot to the entropy.
In a state where the entropy is not maximal there is no reason to find this
infra--red  singularity.

Let us make this idea more concrete by a simple
example of a model which exhibits a first order phase transition in
equilibrium. We take the abelian Higgs model with a small scalar coupling
$\lam\ll e^2$. This model has been studied by several authors and we follow
the notation in \cite{Arnold92} where a discussion of the ring resummation also
 can be found. The Lagrangian is given by
\be{ahlag}
	\cL=-\inv{4}F^2+\abs{D_\mu\Phi}^2+\mu^2\abs{\Phi}^2
	-\inv{3!}\lam\abs{\Phi}^4\ ,
\ee
where $D_\mu\Phi=(\pa_\mu-ieA_\mu)\Phi$. As in \cite{Arnold92} we express the
effective potential in terms of $\phi=\sqrt{2}\abs{\Phi}$. In equilibrium, and
using the Landau gauge, the effective thermal masses are
\be{effmass}
\ba{lcll}
	m^2_{1,\eff}(\phi)&=&{\dpsty -\mu^2+\left(\frac{2\lam}{3}+3e^2\right)
	\frac{T^2}{12}+\frac{\lam\phi^2}{2}}
	&,\ {\rm Higgs};\\[2ex]
	m^2_{2,\eff}(\phi)&=&{\dpsty-\mu^2+\left(\frac{2\lam}{3}+3e^2\right)
	\frac{T^2}{12}+\frac{\lam\phi^2}{6}}
	&,\ {\rm Goldstone};\\[2ex]
	\MT^2(\phi)&=&{\dpsty\frac{e^2T^2}{3}+e^2\phi^2}
	&,\ {\rm Longitudinal};\\[2ex]
	\ML^2(\phi)&=&{\dpsty e^2\phi^2}
	&,\ {\rm Transverse,\ 2\ modes.}
\ea
\ee
The effective potential for the Higgs field is then, using the thermal
masses,
\be{Veq}
	V(\phi)=-\frac{\mu^2}{2}+\frac{\lam}{4!}\phi^4+
	\sum_{i=1}^{5}\left[\frac{T^2}{24}m^2_i(\phi)-
	\frac{T}{12\pi}m^3_i(\phi)\right]\ ,
\ee
where $m_i=\{m_{1,\eff},m_{2,\eff},\ML,\MT,\MT\}$ and $\MT$ occurs twice since
there are two transverse modes. We are here assuming that $T$ is much larger
than any of the masses so that an expansion in $m_i/T$ is accurate. At
the temperature where $V^\bis(\phi=0)=0$ there
is a negative cubic term in $\phi$
which gives a barrier at slightly higher temperatures. In order to conclude
that there is a first order phase transition, the non--trivial minimum has to
be
located at such a $\phi$ that the expansion parameter $e^2 T/m_i(\phi)$ is
small. For instance, when $e=0.1\ \lambda=0.001$ there is a clear first order
phase transition.

This analysis can easily
be repeated for the case of a uniform distribution.
The effective masses are the same as in \eq{effmass} after the
replacement $T^2\goto 3f_{0,B} \pmax^2(t)/\pi^2$. All integrations in the
effective potential can be performed explicitly for the uniform distribution
but we expand the final result in powers of $\max/m$.
 The highest power ($\sim\pmax^2m^2$) gives only the ensemble correction to
the masses and the next term is of the same order as the zero temperature
correction. There is no term linear in $\pmax$ and thus no cubic term in $m_i$.
 We find
\bea{Vunif}
	V(\phi)\align = A\phi^2+B\phi^4~~~~~~~~~~~~~~~~~~~~~~~~~~~~~~~~~~~~
	~~~~~~~~~~~~~~~~~~~~~~~~~~~~~~~~~~~~~~~~~~~~~~~~~\non
	&&+\sum_{i=1}^{5}\left[\frac{f_{0,B}\pmax^2}{8\pi^2}m^2_i(\phi)+
	\frac{f_{0,B} m^4_i(\phi)}{64\pi^2}
	\left(1-\hskip-1pt 4\hskip-1pt \ln(\frac{2\pmax}{m_i(\phi)})\right)+
	\frac{m^4_i(\phi)}{64\pi^2}\ln\frac{m^2_i(\phi)}{2\mu^2}\right]\;.
\eea
The
constants $A$ and $B$ have to be determined by some renormalization condition.

In \eq{Vunif} there is no cubic term in $\phi$, as can be found by expanding
the masses around $\phi=0$, so we do not expect to get the same kind of barrier
 as in equilibrium.
For example, numerical calculations show that with $e=0.1$ there is no
barrier for $\lambda=0.01$ and 0.001. For $\lambda=0.0001$ we do find a small
barrier but it is not accurately described in the perturbation expansion since
$e^2\pmax^2/m_i(\phi)$ is rather large. We thus conclude that in the
absence of an
infra--red singularity, characteristic for the Bose-Einstein distribution,
the abelian
Higgs model has a second order or a very weak first order phase transition.

We may estimate whether the temperature for the phase transition, given by
$m^2_{1,\rm eff}\simeq 0$, is smaller or larger than $T_{th}$ given by \eq{17}.
In fact this can be done for the generic SU(N)--case using Eq.\rf{4}
and Eq.\rf{161}. (Although the latter is strictly speaking true
only well above the phase transition, we may still safely use it
 for an order of
magnitude estimate.) We then find, provided $T_{th}$ is lower
than the equilibration temperature with a thermal background,
 a non--equilibrium phase
transition if $\mu\gg 3.1\times 10^{-4}f_*^{-1/2}f_{0,B}\MP$, irrespective
of the details of the model. Otherwise non--equilibrium phase transition
is always found if $\mu^2\gg C_H^{eq}T^2_{th}$,  and has to be computed
separately for each model. This result is  depicted
in the Figure as a function of the initial particle number.

\Section{concl}{Conclusions}
We have shown that it is rather straightforward
to compute ensemble averages using a  general one--particle
distribution function, assuming statistical independence of the particles.
Many quantities, such as the mass correction, are not very sensitive to the
exact form of the distribution function. They have a typical dependence on
dimensionful parameters and only the
prefactors differ. Other quantities, such as the effective
potential for gauge theories, depend in a more detailed way of
the distribution. The soft part of the Bose-Einstein distribution is
responsible for the first order nature of the phase transition in a
spontaneously broken gauge theory. For the uniform distribution we
find instead a second order phase transition. We have checked that similar
results are found also
for a well localized distribution (a 'delta--peak' distribution).

We have argued that the full thermalization of the uniform distribution takes
place relatively late partially due to a kinematical
suppression of the one--loop
imaginary parts. Similar suppression would arise also in the case of
a delta--peak distribution, but it is not a generic feature. In any case,
for an initial universe out of equilibrium,
the thermalization temperature cannot exceed the limit decreed by
the thermalization
rate in a thermal background, about $5\times 10^{14}$ GeV,
with a weak dependence on the number of degrees of freedom actually present.
Thus
a typical GUT phase transition might have occurred in a non--equilibrium
environment. Such a phase transition would change the particle distribution,
but because the perturbative calculation of $\gamma$ breaks down at the
phase transition, we cannot predict the resulting distribution. However,
it cannot relax to full thermal equilibrium above $T^{eq}_{th}$.
These considerations are relevant for GUT baryogenesis and possibly also
for the
initial conditions in the inflationary
universe models.
\newpage
%

\end{document}